# Vertical Field Effect Transistor based on Graphene-WS$_2$ Heterostructures for flexible and transparent electronics

*Thanasis Georgiou[1], Rashid Jalil[2], Branson D. Belle[2], Liam Britnell[1], Roman V. Gorbachev[2], Sergey V. Morozov[3], Yong-Jin Kim[1,4], Ali Gholinia[5], Sarah J. Haigh[5], Oleg Makarovsky[6], Laurence Eaves[1,6], Leonid A. Ponomarenko[1], Andre K. Geim[2], Kostya S. Novoselov[1], Artem Mishchenko[1]\**

[1]School of Physics and Astronomy, University of Manchester, Manchester M13 9PL, UK.
[2]Manchester Centre for Mesoscience and Nanotechnology, University of Manchester, Manchester M13 9PL, UK.
[3]Institute for Microelectronics Technology RAS, Chernogolovka, 142432, Russia.
[4]Department of Chemistry, College of Natural Sciences, Seoul National University, Seoul 151-747, Korea.
[5]Materials Science Centre, University of Manchester, Manchester M1 7HS, UK.
[6]School of Physics and Astronomy, University of Nottingham, Nottingham NG7 2RD, UK.
*To whom correspondence should be addressed. E-mail: artem.mishchenko@gmail.com

**The celebrated electronic properties of graphene[1,2] have opened way for materials just one-atom-thick[3] to be used in the post-silicon electronic era[4]. An important milestone was the creation of heterostructures based on graphene and other two-dimensional (2D) crystals, which can be assembled in 3D stacks with atomic layer precision[5–7]. These layered structures have already led to a range of fascinating physical phenomena[8–11], and also have been used in demonstrating a prototype field effect tunnelling transistor[12] – a candidate for post-CMOS technology. The range of possible materials which could be incorporated into such stacks is very large. Indeed, there are many other materials where layers are linked by weak van der Waals forces, which can be exfoliated[3,13] and combined together to create novel highly-tailored heterostructures. Here we describe a new generation of field effect vertical tunnelling transistors where 2D tungsten disulphide serves as an atomically thin barrier between two layers of either mechanically exfoliated or CVD-grown graphene. Our devices have unprecedented current modulation exceeding one million at room temperature and can also operate on transparent and flexible substrates.**

Operation of a graphene-based field effect tunnelling transistor (FETT) relies on the tunability of the effective height of the tunnelling barrier ($\Delta$) via the shift of the Fermi level in graphene and changes in the height and shape of the barrier. Previously, devices with barriers formed by thin layers of hexagonal boron nitride[14] (hBN) or the silicon/graphene Schottky interface[15] were considered. In this letter, we demonstrate that by utilising both the tunnelling and thermionic over-barrier currents in a tungsten disulphide (WS$_2$)-based FETT, a dramatic improvement of the device characteristics can be achieved, so that they satisfy the requirements for next generation electronic devices. Bulk WS$_2$ is an indirect band gap (1.4 eV) semiconductor, which is expected to turn into a direct gap (2.1 eV) material when exfoliated into the monolayer state[16]. Each single plane of WS$_2$ consists of a trilayer: a tungsten layer sandwiched by two sulphur layers in a trigonal prismatic coordination. Using barrier materials with a relatively small band gap (such as WS$_2$) provides a viable method to increase the ON/OFF ratio of the FETT, since the changes in the Fermi level of graphene (usually < 0.5 eV as limited by the gate dielectric breakdown) are of the order of, or even exceed the barrier height. In contrast, hBN is a wide-gap insulator (E$_g$>5 eV) and forms a high tunnelling barrier in the graphene-hBN FETT, where the changes in the Fermi level of graphene are small compared to the barrier height. Being chemically stable and having only a weak impurity band, WS$_2$ also offers a distinct advantage over molybdenum disulphide (MoS$_2$)[12,17], since it allows switching between tunnelling and thermionic transport regime, resulting in much better transistor characteristics and thus allowing for much higher ON/OFF ratios and much larger ON current.

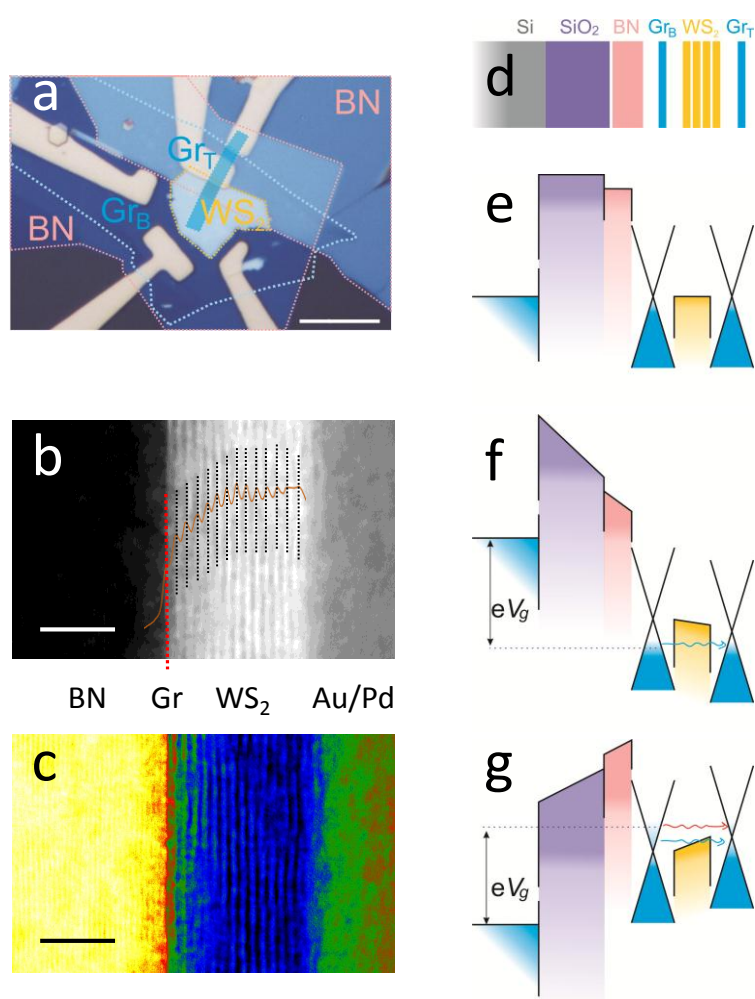

**Figure 1.** Graphene-WS$_2$ heterotransistor. **a,** Optical image (scale bar 10μm), **b,** Cross-section high-resolution HAADF STEM image (scale bar 5nm) and **c,** Bright field STEM image (scale bar 5nm). **d,** Schematic of transistor vertical architecture. **e,** Band diagram corresponding to no $V_g$, $V_b$ being applied. **f,** A negative $V_g$ shifts the Fermi level of the two graphene layers down from the neutrality point, increasing the potential barrier and switching the transistor OFF. **g,** Applying a positive $V_g$ results in an increased current to flow between Gr$_B$ and Gr$_T$ due to both thermionic (red arrow) and tunnelling (blue arrow) contributions.

Figure 1 shows schematic and band-diagrams of our devices. The fabrication techniques are detailed in the Methods section. Briefly, a dry transfer procedure was used to assemble heterostructures from individual flakes prepared by micromechanical exfoliation. The first layer of graphene (bottom contact) placed on hBN was covered by a thin WS$_2$ flake. We used graphene (mechanically exfoliated or CVD-grown), few layer graphene, graphite or Cr/Au as top contacts. More than a dozen graphene-WS$_2$ heterostructures have been studied and qualitatively similar behaviour was observed, regardless of the material used as a top contact. For simplicity, to illustrate the operation of such devices, we discuss transistors where both bottom and top contacts where made of graphene layers. It is worth mentioning that the highest ON/OFF ratio was found with 4-5 layer-thick $WS_2$ tunnelling barriers.

The principle of operation of $WS_2$-based FETT is illustrated by the series of band diagrams presented in Fig. 1. Without gate or bias voltage applied, the Fermi levels of the two graphene sheets are located near the bottom of the conduction band of $WS_2$, Fig. 1e (as can be seen from temperature dependence below). Gate voltage $V_g$ applied between the silicon substrate and the bottom graphene layer, $Gr_B$, changes the carrier concentration $n$ in $Gr_B$, shifting the Fermi level by $\Delta E_F = \pm \hbar v_F \sqrt{\pi |n|}$. The sign of the Fermi level shift is determined by the polarity of the gate voltage. Fig. 1f shows that negative gate voltage shifts $E_F$ down and therefore increases the tunnelling barrier by $\Delta E_F$, giving the "OFF" state of the transistor. A positive gate voltage shifts the Fermi level into the conduction band, reducing the tunnelling barrier and even results in an over-barrier thermionic current (Fig. 1g), thus representing the "ON" state. In addition, because of the weak screening by monolayer graphene, the electric field from the gate electrode penetrates through $Gr_B$, and both changes the shape of the barrier and induces charge carriers in the top graphene layer, $Gr_T$, Fig. 1f,g.

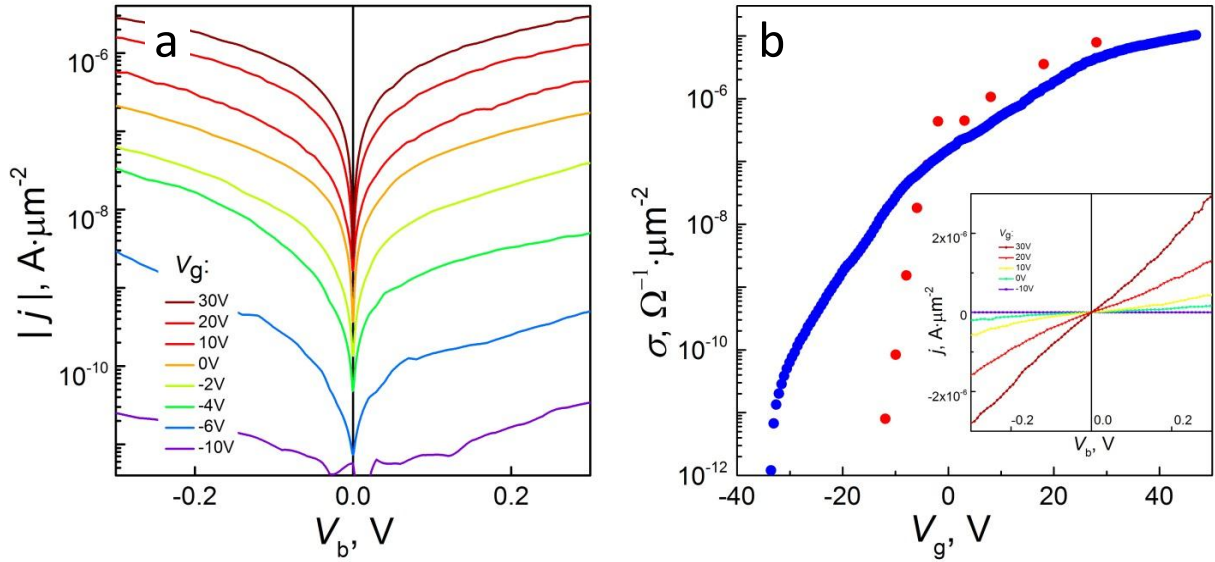

**Figure 2.** Room temperature tunnelling transport measurements in our graphene-$WS_2$ transistor. **a,** $I$-$V_b$ curves for different $V_g$, semi-log scale. **b,** Red circles: zero-bias conductivity as a function of gate voltage (measured from the slope of $I$-$V_b$ curve at zero $V_b$). Blue circles: conductivity measured at $V_b$ = 20 mV as a function of gate voltage. Inset: $I$-$V_b$ curves at different $V_g$, linear scale. T = 300K. Active device area 0.25 $\mu m^2$.

Let us first discuss the case where $V_g$=0 V. At low temperatures and low bias voltages $V_b$, the tunnelling current response is linear with $V_b$, yielding typical resistivity of ≈2-4 M$\Omega\cdot\mu m^2$ for a 4-5 layer thick $WS_2$ barrier at room temperature. At higher biases the current grows exponentially with $V_b$, before showing signatures of saturation (for log($I$)-$V_b$ plot, see Fig. 2a) at $V_b > 50$mV. For $V_g < -10$V, the zero-bias resistivity increases to ≈0.5-1 T$\Omega\cdot\mu m^2$, a value limited by the leakage current through the gate dielectric and remnant wafer contamination. By a applying positive gate voltage we were able to decrease resistivity down to ≈100-200 k$\Omega\cdot\mu m^2$. To reveal transistor operation, zero-bias conductivity is plotted as a function of the gate voltage in Fig. 2b. In contrast to the use of hBN as a barrier, this dependence is no longer dominated by the linear contribution of graphene's density of states. Instead, it behaves exponentially, allowing us to achieve an ON/OFF ratio exceeding $10^6$ even at room temperature.

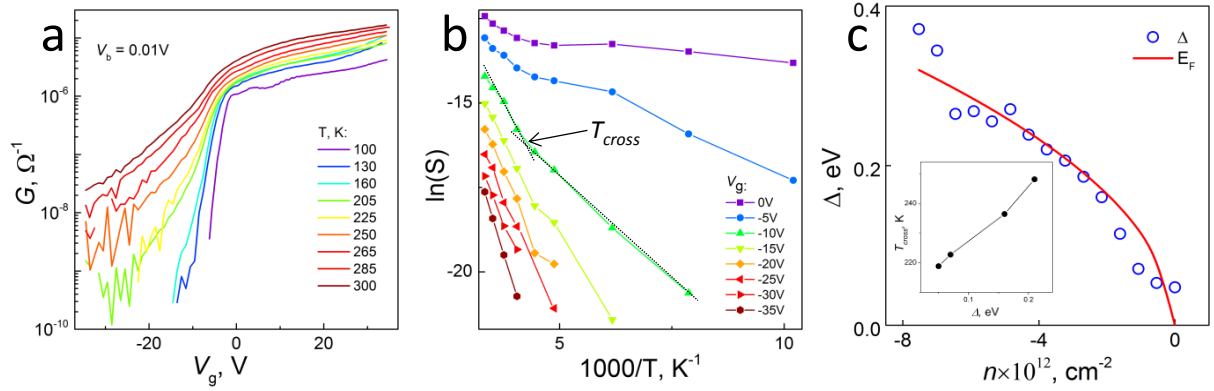

**Figure 3.** Temperature-dependent transistor characteristics. **a,** Conductance – gate voltage plot for T = 100–300K. **b,** Arrhenius plot of small-bias conductance at different gate voltages. **c,** Barrier height $\Delta$, against carrier concentration, $n$, in $Gr_B$ layer (blue circles), calculated Fermi level of graphene (red line). Inset: Measured temperature of transition between thermionic and tunnelling contributions versus barrier height.

In order to obtain a better insight into the mechanism of transport through the $WS_2$ barrier, we study the temperature dependence of the tunnelling current at various $V_g$ and $V_b$. (Fig. 3a,b). Fig. 3b clearly shows a crossover, $T_{cross}$, between the high temperature regime (for instance $T_{cross} \approx 230$ K for $V_g = -10$V) where an exponential temperature dependence is observed and a low temperature regime with a very weak temperature dependence ($T < 240$ K for $V_g = -10$V). Note, that for gate voltages below -15 V the current sensitivity does not allow one to see the crossover.

The two temperature regimes indicate two different mechanisms for electron transport through the barrier, Fig. 1g. At low temperatures or very large negative gate voltages the under-barrier electron tunnelling prevails. The tunnelling current can be described in terms of the available tunnelling density of states in the two graphene contacts, $DoS(E) \cdot f(E)$ (here $f(E) = \dfrac{1}{e^{\Delta E/k_B T} + 1}$ is the Fermi-Dirac distribution) and the probability $T(E)$ that electron penetrates the barrier[11,12,18]:

$$I(V) \propto \int dE \cdot DoS_B(E) \cdot DoS_T(E-eV) \cdot \left[ f(E-eV) - f(E) \right] \cdot T(E)$$

For a parabolic band semiconductor the transmission coefficient is $T(E) \approx e^{\frac{-2w\sqrt{2m^*\Delta}}{\hbar}}$, where $m^*$ is the effective mass of electron in the semiconductor, $w$ and $\Delta$ are the barrier width and height, respectively. As was shown previously, this standard model is well applicable to layered crystals[12]. In this regime the tunnelling current scales exponentially with the thickness of the barrier.

At high temperatures and positive gate voltages the transport is dominated by the over-barrier thermionic current. The number of thermally excited carriers with energy above the barrier can be estimated from the Boltzmann distribution, $f(E) \approx e^{-\Delta/k_B T}$. For small enough barriers, current will also be exponentially sensitive to the reciprocal temperature. Thus, we used the high temperature part of the zero-bias conductivity dependence to extract $\Delta$, as a function of carrier concentration, $n$. The observed square-root dependence $\Delta \propto \sqrt{|n|}$ (Fig. 3c) clearly indicates that at zero gate voltage the Fermi level in $Gr_b$ is located near the Dirac point and aligned close to the bottom of the conduction band in $WS_2$. It can be tuned electrostatically by the external gate which determines the change in the barrier height. The thermionic transport mechanism lifts the limit of the ON current, which usually exists in tunnelling transistors.

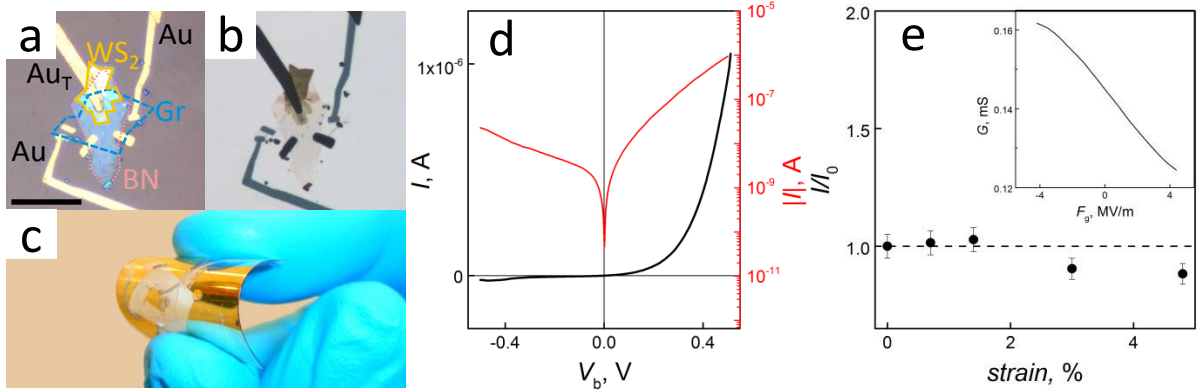

**Figure 4.** Transistor operation on flexible and transparent substrates. **a,b,** Optical images of a device in reflection and transmission modes, respectively. Scale bar 10 μm. **c,** Image of the device under bending. **d,** IV of the same device with $V_g$=0V and bending applied. Curvature is 0.05 mm$^{-1}$ and T = 300 K. **e,** Relative current variation vs. applied strain. Inset: gating transistor under strain.

Beyond its peculiar electrical properties graphene also has good optical[19,20] and mechanical[21] properties. Other 2D materials are also naturally transparent[3,14,22] and often demonstrate a significant Young modulus and high breaking strength[23,24] (as for $MoS_2$, $WS_2$ is expected to demonstrate similar characteristics). Thus, it is reasonable to suggest that 2D-based heterostructures can be attractive for transparent and flexible electronics. Indeed, our transistors are only a few atomic layers thick and the device should not experience any bending-induced effects due to its vertical geometry. To this end we prepared such vertical heterostructures on flexible PET films (Fig. 4a-c) using the same transferring procedure (just avoiding annealing step, see Methods). The tunnelling characteristics achieved (Fig. 4d) are very similar to those obtained for devices on $SiO_2$, and practically do not change with bending (Fig. 4e). Our flexible structures also demonstrate transistor effect (Fig. 4e, inset), albeit small due to a large distance between the tunnelling junction and the gate electrode (which was deposited on the opposite surface of the PET film). The effect demonstrates the feasibility of the concept and calls for future optimisation.

In conclusion, we have demonstrated that $WS_2$ is a material which acts as an ideal vertical transport barrier in graphene heterostructures. The achieved ON/OFF ratio is among the highest found in the literature for a graphene-based transistor device. The vertical field effect transistors also offer possibilities for flexible and transparent electronics, since for operation the FETT is only required to be few atoms thick. Moreover, we demonstrate device operation from CVD-grown materials, an imperative scalability requirement.

# Acknowledgements

This work was supported by the European Research Council, European Commission FP7, Engineering and Physical Research Council (UK), the Royal Society, U.S. Office of Naval Research, U.S. Air Force Office of Scientific Research, and the Körber Foundation. A.M. acknowledges support from the Swiss National Science Foundation. Y.-J.K. was supported by the Global Research Laboratory (GRL) Program (2011-0021972) of the Ministry of Education, Science and Technology, Korea.

# Author contributions

T.G., R.J., B.D.B. and R.V.G. fabricated the devices. Y.J.K. grew the CVD graphene. A.G. and S.J.H. did STEM imaging. O.M. and L.E. designed the setup for low-current measurements. L.B., S.V.M. and A.M. performed transport measurements. L.A.P., A.K.G., K.S.N and A.M. conceived and designed the experiments. T.G., A.M. and K.S.N. wrote the manuscript. All authors made critical contributions to the work, discussed the results and commented on the manuscript.

# Methods

**Device fabrication**

Our graphene–$WS_2$ heterostructures are prepared using techniques described previously. First, a thick (~30 nm) flake of a hexagonal boron nitride was mechanically exfoliated on a degenerately doped Si substrate covered with a thin layer (either 90 or 290 nm) of thermally grown $SiO_2$. The Si substrate also serves as the system's global back gate. Subsequent flakes were transferred by a so-called dry transfer technique[6,8]. The technique is based on the use of a dual-polymer substrate: flakes are deposited on the top polymer and by dissolving the bottom one; we release the top polymer-flake bilayer, which is then carefully inverted and aligned on the target flake with micrometre accuracy.

The dry transfer technique is repeated four times to prepare the devices described in main text. First, transfer on the boron nitride substrate a layer of graphene, $Gr_B$, (annealing in forming gas 250°C), second, a few layer flake of $WS_2$, third, a top BN layer to isolate $WS_2$ from the bottom graphene electrode and finally a top contact of graphene or graphite on $WS_2$, $Gr_T$. The graphene and few-layer graphene flakes are subsequently independently contacted following standard microfabrication and metallization techniques (3nmCr/50nmAu). In a simpler version of the device, the top electrode can be a metal, reducing the number of fabrication steps.

**Transport measurements**

I-V characteristics of graphene-based heterostructure FETs were measured as a function of both gate voltage and temperature (300K – 100K) in a helium atmosphere with a Keithley 2636A SourceMeter.

**Scanning Transmission Electron Microscopy**

Scanning transmission electron microscope (STEM) imaging was carried out using a Titan G2 probe-side aberration corrected STEM operated at 200 kV and equipped with a high efficiency ChemiSTEM energy dispersive x-ray detector. The convergence angle was 19 mrad and the third-order spherical aberration was set to zero (±5 μm). The multilayer structures were oriented along an ⟨hkl0⟩ crystallographic direction by taking advantage of the Kikuchi bands of the Si substrate.

**Graphene CVD growth**

Large-area, high-quality graphene films were synthesized on copper (Cu) foil using chemical vapor deposition (CVD). First, 25μm-thick Cu foil (Alfa Aesar, item No. 13382) was cleaned with acetone, deionized water and isopropanol. The Cu foil was then loaded into a quartz tube CVD chamber, evacuated and heated to 1000°C with an $H_2$ flow at 20 standard cubic centimetres per minute (sccm) at 200mTorr. The Cu foil was annealed at 1000°C for 30 min to increase the Cu grain size and to remove the native oxide layer from the surface. For growth of a continuous graphene film, a gas mixture of $CH_4$ and $H_2$ at flow rates of 40 and 20 sccm were introduced into the CVD chamber for 30 min. During growth, the reactor pressure and temperature were maintained at 1000°C and 600mTorr, respectively. Finally, the sample was cooled rapidly to room temperature in a hydrogen atmosphere at a pressure of 200mTorr. The details of the graphene preparation and transfer process are described elsewhere[25].

# References

1. Novoselov, K. S. *et al.* Electric field effect in atomically thin carbon films. *Science* **306**, 666–669 (2004).
2. Castro Neto, A. H., Guinea, F., Peres, N. M. R., Novoselov, K. S. & Geim, A. K. The electronic properties of graphene. *Rev. Mod Phys.* **81**, 109–162 (2009).
3. Novoselov, K. S. *et al.* Two-dimensional atomic crystals. *Proc. Natl. Acad. Sci. USA.* **102**, 10451–10453 (2005).
4. Geim, A. K. Graphene: status and prospects. *Science* **324**, 1530–1534 (2009).
5. Haigh, S. J. *et al.* Cross-sectional imaging of individual layers and buried interfaces of graphene-based heterostructures and superlattices. *Nature Mater.* **11**, 1–4 (2012).
6. Dean, C. R. *et al.* Boron nitride substrates for high-quality graphene electronics. *Nature Nanotech.* **5**, 722–726 (2010).
7. Novoselov, K. S. Nobel Lecture: Graphene: Materials in the Flatland. *Rev. Mod. Phys.* **83**, 837–849 (2011).
8. Ponomarenko, L. A. *et al.* Tunable metal–insulator transition in double-layer graphene heterostructures. *Nature Phys.* **7**, 958–961 (2011).
9. Gorbachev, R. V. *et al.* Strong Coulomb drag and broken symmetry in double-layer graphene. *Nature Phys.* (2012) *advance online publication,* doi: 10.1038/nphys2441.
10. Kim, S. *et al.* Coulomb drag of massless fermions in graphene. *Phys. Rev. B* **83**, 161401 (2011).
11. Britnell, L. *et al.* Electron tunneling through ultrathin boron nitride crystalline barriers. *Nano Lett.* **12**, 1707–1710 (2012).
12. Britnell, L. *et al.* Field-effect tunneling transistor based on vertical graphene heterostructures. *Science* **335**, 947–950 (2012).
13. Coleman, J. N. *et al.* Two-dimensional nanosheets produced by liquid exfoliation of layered materials. *Science*  **331**, 568–571 (2011).
14. Gorbachev, R. V. *et al.* Hunting for monolayer boron nitride: optical and Raman signatures. *Small* **7**, 465–468 (2011).
15. Yang, H. *et al.* Graphene barristor, a triode device with a gate-controlled Schottky barrier. *Science* **336**, 1140–1143 (2012).
16. Kuc, A., Zibouche, N. & Heine, T. Influence of quantum confinement on the electronic structure of the transition metal sulfide $TS_2$. *Phys. Rev. B* **83**, 245213 (2011).

17. Sliney, H. E. Solid lubricant materials for high temperatures—a review. *Tribol. Int.* **15**, 303–315 (1982).
18. Simmons, J. G. Generalized Formula for the Electric Tunnel Effect between Similar Electrodes Separated by a Thin Insulating Film. *J. Appl. Phys.* **34**, 1793–1803 (1963).
19. Nair, R. R. *et al.* Fine structure constant defines visual transparency of graphene. *Science* **320**, 1308 (2008).
20. Blake, P. *et al.* Making graphene visible. *Appl. Phys. Lett.* **91**, 063124 (2007).
21. Lee, C., Wei, X., Kysar, J. W. & Hone, J. Measurement of the elastic properties and intrinsic strength of monolayer graphene. *Science* **321**, 385–388 (2008).
22. Castellanos-Gomez, A., Agraït, N. & Rubio-Bollinger, G. Optical identification of atomically thin dichalcogenide crystals. *Appl. Phys. Lett.* **96**, 213116 (2010).
23. Bertolazzi, S., Brivio, J. & Kis, A. Stretching and breaking of ultrathin $MoS_2$. *ACS Nano* **5**, 9703–9709 (2011).
24. Andrew, R., Mapasha, R., Ukpong, A. & Chetty, N. Mechanical properties of graphene and boronitrene. *Phys. Rev. B* **85**, 125428 (2012).
25. Bae, S. *et al.* Roll-to-roll production of 30-inch graphene films for transparent electrodes. *Nature Nanotech.* **5**, 574–578 (2010).

# Supplementary information:

# Vertical Field Effect Transistor based on Graphene-$WS_2$ Heterostructures for flexible and transparent electronics

1. **Device characterization: AFM imaging, Raman Spectroscopy and cross-sectional STEM imaging.**

Topography images were obtained with AFM Nanoscope Dimension V (Bruker) in tapping mode (PPP-NCHR from Nanosensors) under ambient conditions. Figure S1 shows the high quality and uniformity of the fabricated graphene-$WS_2$ heterojunction.

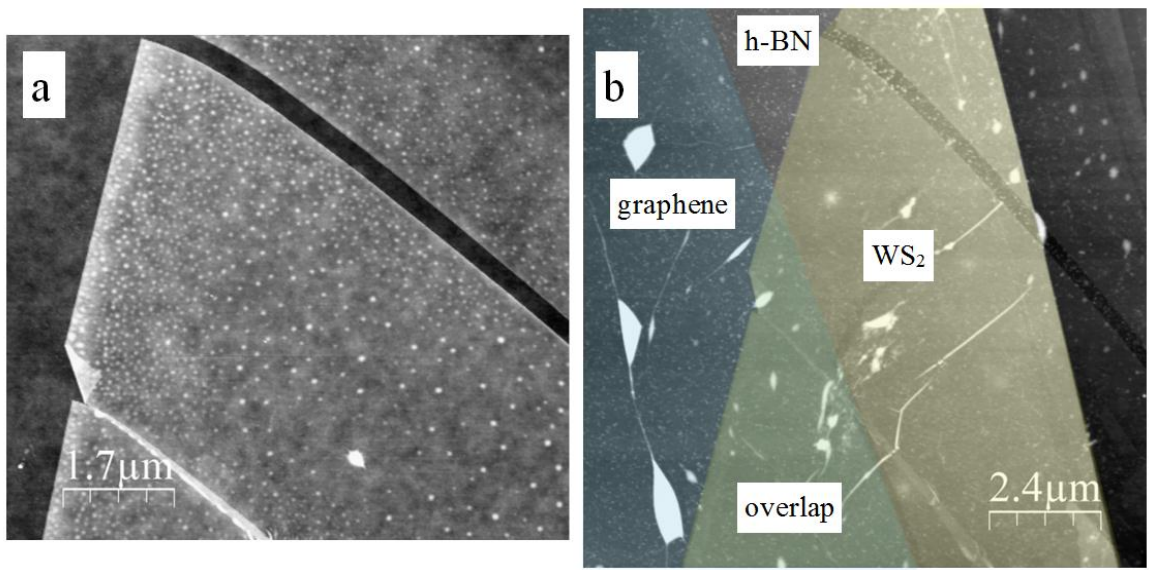

Figure S1. AFM image of $WS_2$ trilayer before (a) and after (b) transfer to graphene.

**Raman Spectroscopy**

Raman spectroscopy has proven to be a fast, non-destructive technique for characterizing two-dimensional materials. Layered TDMCs are all expected to share two main features in their Raman spectrum: an in-plane $E_{2g}$ mode where the metal and chalcogen atoms vibrate in opposite directions and an out-of-plane $A_{1g}$ mode where the chalcogen atoms vibrate out-of-plane. For bulk $WS_2$, the $E_{2g}$ and $A_{1g}$ modes lie at 352cm$^{-1}$ and 421cm$^{-1}$, respectively. The transition from bulk to single atomic layer is captured in the Raman spectrum. Figure S2 shows the Raman spectrum of $WS_2$ for a single, bi- and multi-layer flake captured by a Renishaw spectrometer with 2.4eV excitation line. The $E_{2g}$ mode is found to red shift by 2cm$^{-1}$ while the $A_{1g}$ mode blue shifts by also 2cm$^{-1}$. The shifts are similar to those previously observed in atomically thin $MoS_2$ [S1, S2]. The red shift of the $E_{2g}$ mode is due to a reduction of the dielectric screening of a monolayer compared to the bulk [S3], while the blue-shift of the $A_{1g}$ mode is due to the decrease of the restoring force acting on the atoms as a result of the decreased thickness [S1].

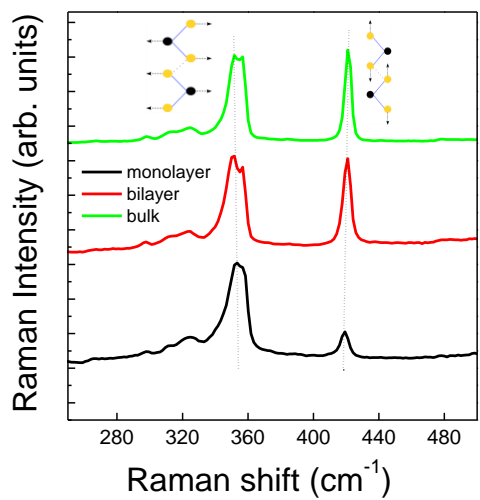

Figure S2. Thickness-dependent Raman spectroscopy of atomically thin $WS_2$. The $E_{2g}$ mode softens while the $A_{1g}$ hardens.

Probing the photoluminescence of $WS_2$, we also observe the transition to a direct-gap semiconductor when in a monolayer state. This is seen as a large increase in the intensity of the photoluminescence compared to its bilayer or bulk, Figure S3.

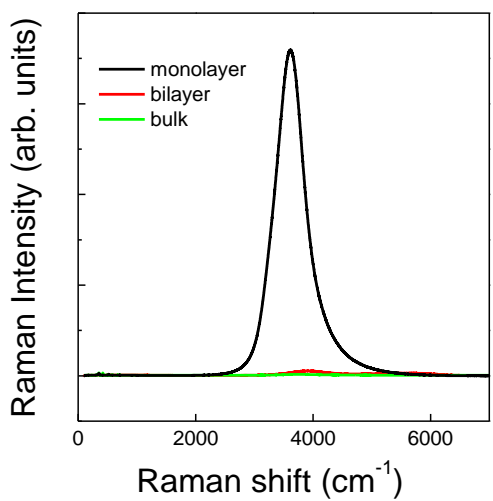

Figure S3. Photoluminescence spectroscopy for $WS_2$ of different thickness, normalized with respect to the Raman $E_{2g}$ mode.

**Preparation of TEM samples and cross-sectional STEM imaging**

A dual-beam instrument (FEI Nova NanoLab 600) has been used for site-specific preparation of cross-sectional samples suitable for TEM analysis using the lift-out approach. This instrument combines a FIB and a SEM column in the same chamber and is also fitted with a gas-injection system to allow local material deposition and material-specific preferential milling to be performed by introducing reactive gases in the vicinity of the electron or ion probe. The electron column delivers the imaging abilities of the SEM and is at the same time less destructive than FIB imaging, Figure S4. SEM imaging of the device before milling allows one to identify an area suitable for side-view imaging. After sputtering of a 50nm Au–Pd coating on the whole surface ex situ, the Au/Ti contacts on graphene were still visible as raised regions in the secondary electron image. These were used to correctly position the ion beam so that a Pt strap layer could be deposited on the surface at a chosen location, increasing the metallic layer above the device to ~1μm. The strap protects the region of interest during milling as well as providing mechanical stability to the cross-sectional slice after its removal. Trenches were milled around the strap by using a 30kV $Ga^+$ beam with a current of 0.1–5nA. Before removing the final edge supporting the milled slice and milling beneath it to free it from the substrate, one end of the Pt strap slice was welded to a nanomanipulator needle using further Pt deposition. The cross-sectional slice with typical dimensions of 1μm × 5μm × 10μm could then be extracted and transferred to an Omniprobe copper half grid, as required for TEM. The slice was then welded onto the grid using further Pt deposition so that it could be safely separated from the nanomanipulator by FIB milling. A final gentle polish with $Ga^+$ ions (at 5kV and 50pA) was used to remove side damage and reduce the specimen thickness to less than 50nm. The fact that the cross-sectional slice was precisely extracted from the chosen spot was confirmed by comparing the position of identifiable features such as Au contacts and termination of the $WS_2$ layer, which are visible both in the SEM images of the original device and within TEM images of the prepared cross-section.

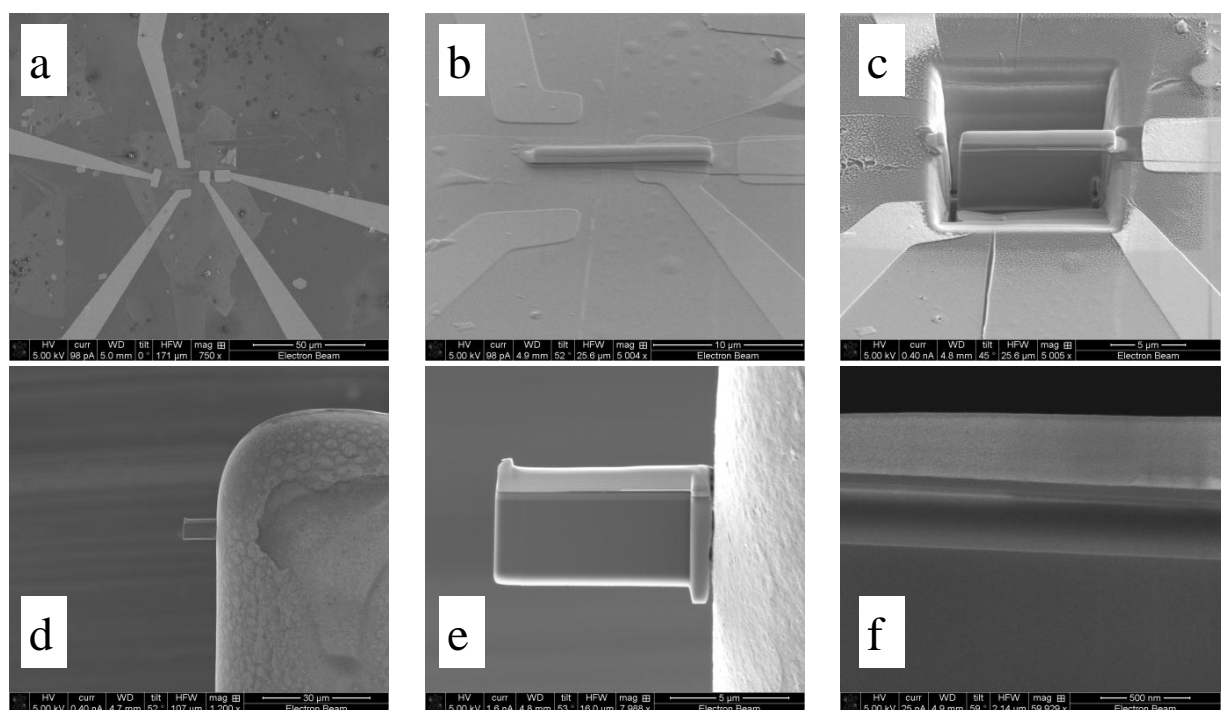

Figure. S4. Sample preparation for cross-sectional STEM imaging. **a,** SEM of the device, **b,** Pt strap layer deposited, **c,** FEB milling of the device, **d,e,** slice for the cross-sectional TEM, **f,** SEM image of the cross-section of the prepared specimen.

## 2. FETT performance: barrier thickness

Since tunnelling current decays exponentially with the thickness of the barrier, transport in devices with thick barriers will be dominated by diffusion current through $WS_2$. In this case the transistor would operate as a very short channel FET. Figure S5 shows the FETT with thick-$WS_2$ barrier operated at two different temperatures. Carrier freeze-out at lower temperature indicates that diffusive transport through thick $WS_2$ is predominant.

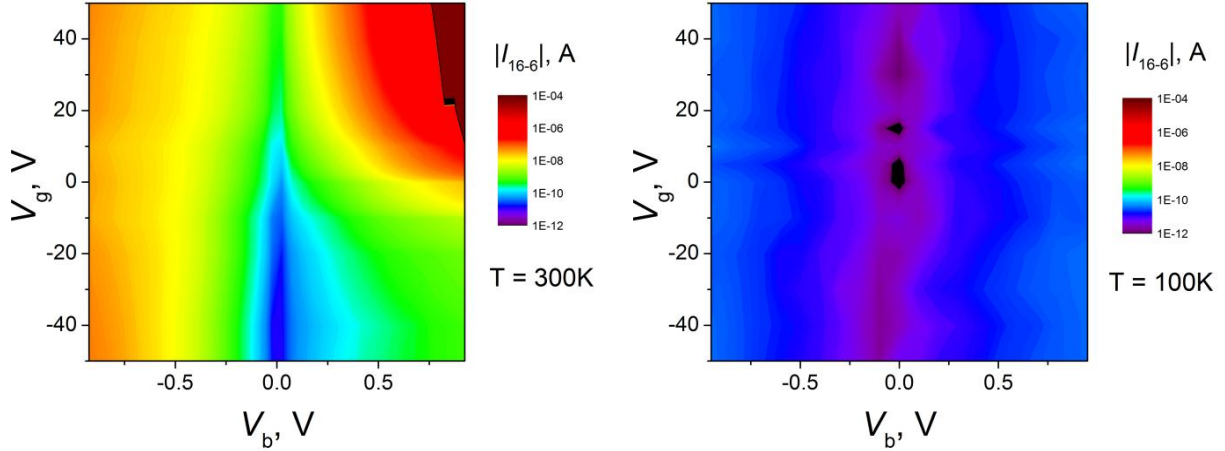

Figure S5. Current map $I = f(V_b, V_g)$ at T = 300K and 100K for a FETT with thick (>20nm) $WS_2$ layer.

For thin $WS_2$ barriers (3-8 layers) we did not observe such freeze-out, see e.g. Fig. 3a in the main text. In case of mono- and bilayer $WS_2$ the OFF current becomes unacceptably high due to the wavefunction penetration through the ultrathin barrier.

## 3. FETT performance: temperature dependence

In order to visualize the interplay between thermionic and tunnelling contributions to the total current in FETT we model these two contributions. The tunnelling part can be estimated as an integral

$$I(V) \propto \int_{-\infty}^{\Delta} dE \cdot DoS_B(E) \cdot DoS_T(E-eV) \cdot \left[ f(E-eV) - f(E) \right] \cdot T ,$$

where $DoS(E) = \dfrac{2|E|}{\pi \hbar^2 v_F^2}$, $T = \text{Exp}\left( \dfrac{-2\sqrt{2m^*}}{\hbar} \int_0^{d_{barrier}} dx \sqrt{\Delta - \dfrac{V \cdot x}{d_{barrier}}} \right)$ and $f(E) = \dfrac{1}{e^{(E-\mu)/k_B T} + 1}$

The thermionic current can be estimated by taking the integral from $\Delta$ to $+\infty$ and assuming $T = 1$:

$$I(V) \propto \int_{\Delta}^{+\infty} dE \cdot DoS_B(E) \cdot DoS_T(E-eV) \cdot \left[ f(E-eV) - f(E) \right] \cdot 1$$

Results of this simulation presented on Figure S6 demonstrate the interplay between tunnelling and thermionic currents at three different temperatures.

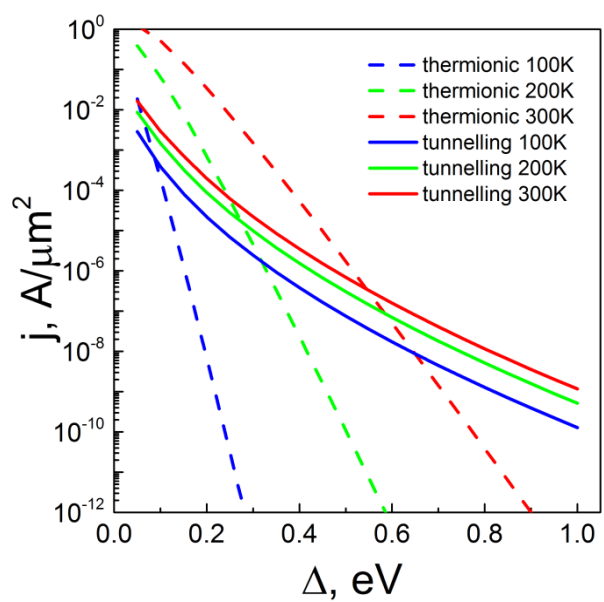

Figure S6. Simulation of tunnelling and thermionic contributions to the total current as a function of the barrier height at three different temperatures: 100K, 200K and 300K.

**References:**

S1.  Lee, C. *et al.* Anomalous lattice vibrations of single- and few-layer $MoS_2$. *ACS Nano* **4**, 2695–2700 (2010).

S2.  Li, H. *et al.* From Bulk to Monolayer $MoS_2$: Evolution of Raman Scattering. *Adv. Funct. Mater.* (2012).

S3.  Molina-Sánchez, a. & Wirtz, L. Phonons in single-layer and few-layer $MoS_2$ and $WS_2$. *Phys. Rev. B* **84**, 155413 (2011).

S4.  The International Technology Roadmap for Semiconductors. http://www.itrs. net/Links/2009ITRS/Home2009.htm (2009).

S5.  Lundstrom, M. Moore's Law Forever? *Science* **773**, 210–211 (2003).

S6.  Zhang, Q., Zhao, W. & Seabaugh, A. Low-subthreshold-swing tunnel transistors. *IEEE Electr. Device. L.* **27**, 297–300 (2006).

S7.  Bhuwalka, K. K., Schulze, J. & Eisele, I. Performance Enhancement of Vertical Tunnel Field-Effect Transistor with SiGe in the δp + Layer. *Jpn. J. Appl. Phys.* **43**, 4073–4078 (2004).